\documentclass[aps,twocolumn]{revtex4-1}
\usepackage{amsmath}
\usepackage{bm}
\usepackage{graphicx}

\begin{document}

\title{Transdimensional epsilon-near-zero modes in planar plasmonic nanostructures}

\author{Igor V. Bondarev}
\email[Corresponding author email:~]{ibondarev@nccu.edu}
\affiliation{Department of Math \& Physics, North Carolina Central University, Durham, NC 27707, USA}
\author{Hamze Mousavi}
\affiliation{Department of Math \& Physics, North Carolina Central University, Durham, NC 27707, USA}
\author{Vladimir M. Shalaev}
\affiliation{School of Electrical \& Computer Engineering and Birck Nanotechnology Center,
Purdue University, West Lafayette, IN 47907, USA}

\begin{abstract}
We use quantum electrodynamics and the confinement-induced nonlocal dielectric response model based on the Keldysh-Rytova electron interaction potential to study the epsilon-near-zero modes of metallic films in the transdimensional regime. New peculiar effects are revealed such as the plasmon mode degeneracy lifting and the dipole emitter coupling to the split epsilon-near-zero modes, leading to thickness-controlled spontaneous decay with up to three-orders-of-magnitude increased rates.
\end{abstract}

\maketitle

\section{Introduction}

Transdimensional (TD) materials are ultrathin planar nano\-structures composed of a precisely controlled finite number of monolayers~\cite{aebshal19}. Modern material fabrication techniques allow one to produce stoichiometrically perfect films of metals and semiconductors down to a few, or even a single monolayer in thickness~\cite{thingold,thinXenes,Shah17,javierOptica19,javierACSN19}. TD mate\-rials make it possible to probe fundamental properties of light-matter interactions as they evolve from a single atomic layer to a larger number of layers approaching the bulk material properties. The current research has been largely focusing on either purely 2D structures including metal-dielectric interfaces and novel 2D materials~\cite{2D1,2D2}, or on conventional bulk materials, being guided by the traditional view that only the dimensionality and chemical composition are important to control the optoelectronic properties of materials. The transitional, transdimensional regime laying in between 3D and 2D, has been largely out of the major research focus so far.

Ultrathin films made of metals, doped semiconductors, or polar materials with a thickness of only a~few atomic layers, can support plasmon-, exciton-, and phonon-polariton eigenmodes~\cite{javierACSN19,2D1,2D2,JoshNL19,Zhelud19,Shah18,Brener15,Greffet19}.~Such TD materials are therefore expected to show the high tailorability of their electronic and optical properties mainly by varying their thickness (number of monolayers) in addition to the standard possibility of altering their chemical, atomic, and electronic composition (stoichiometry, doping).~This makes TD materials distinctly different from conventional thin films commonly described by either purely 2D material properties, or by bulk (3D) materials with boundary conditions imposed on their top and bottom interfaces~\cite{Ritchie57,Economou69,DahlSham77,Theis80,AndoFowlerStern82,Chaplik85,Wang96,Pitarke07,Politano14}. Plasmonic TD materials (ultrathin finite-thickness metallic films) can provide controlled light confinement due to their thickness-dependent localized surface plasmon (SP) modes~\cite{Brener15,Greffet19}, thus offering tunable light-matter coupling, higher adjustable transparency, and new quantum phenomena such as enabling atomic transitions that are normally forbidden~\cite{RiveraSci16}. Similar to truly 2D and quasi-2D materials such as graphene and transition metal dichalcogenide monolayers~\cite{2D2,Basov}, plasmonic TD materials are expected to show the extreme sensitivity to external fields, making possible advances such as novel parity-time symmetry breaking photonic designs~\cite{Engheta13,AbramNL18} that can further develop the fields of nanophotonics, plasmonics, and optical metasurfaces~\cite{BozhevACS18}. However, while some predictions on anomalous dispersion and tunable light confinement in plasmonic TD films are made~\cite{mangar14,garman15,bondshal17,bondmoushal18,bond19}, much remains unclear about their optical response and quantum near-field effects.

Here we use macroscopic quantum electrodynamics (QED) and the confinement-induced \emph{nonlocal} Drude dielectric response model based on the Keldysh-Rytova (KR) pairwise electron interaction potential~\cite{bondshal17,bondmoushal18,bond19}, to study epsilon-near-zero (ENZ) modes and their coupling to a point-like atomic dipole emitter (DE) near the surface of the metallic film in the TD regime. The ENZ modes are vertically confined SP modes of frequency $\omega(k)$ reaching the plasma oscillation frequency $\omega_p$ of the film whereby its dielectric response function crosses zero~\cite{Brener15}. Using the KR model, we have earlier shown the ultrathin TD plasmonic films to exhibit remarkable properties such as the low-frequency negative refraction and the resonance magneto-optical response~\cite{bondmoushal18}.

Here, using the same KR model we report new salient thickness-controlled features for TD plasmonic films: the SP mode degeneracy lifting and DE coupling to the ENZ modes split. This coupling is shown to lead to the biexponential distance dependence of the spontaneous decay with rates raised by two-to-three orders of magnitude as compared to free space. Remarkably, these effects can be controlled due to the thickness-dependent plasma frequency of the film~\cite{bondshal17} --- a unique microscopic property that cannot be obtained from macroscopic boundary conditions on the bulk material film interfaces, the property that originates from the vertical electron confinement to change the electron-electron Coulomb potential into the much stronger KR interaction potential~\cite{Keldysh79}. Our results generalize the fundamental work by Drexhage~\cite{Drexhage} and related recent works~\cite{Brener15,Greffet19} by explicitly taking into account the confinement effects in TD films.

\section{Confinement-induced nonlocality}

The electrostatic Coulomb field produced by remote charges confined in the film outside of their confinement region starts playing a perceptible role with the film thickness reduction. If the environment has a lower dielectric constant than that of the structure confined [as it shows in the inset of Fig.~\ref{fig3}~(a) with $\varepsilon_{1,2}\!<\!\varepsilon$], the increased 'outside' contribution makes the Coulomb interaction of the charges confined much stronger than that in a homogeneous medium with the same dielectric constant. The 3D pairwise electrostatic Coulomb interaction potential takes the thickness-dependent 2D form known as the KR potential~\cite{Keldysh79}. This is a solely confinement-induced effect known to occur both for quasi-2D and for quasi-1D confined geometries~\cite{Chernikov14,Louie09}. As a consequence, the in-plane plasma oscillation frequency in the Drude response of the metallic film of thickness $d$ takes the form~\cite{bondshal17}
\begin{equation}
\omega_p=\omega_p(k)=\frac{\omega_p^{3D}}{\sqrt{1+(\varepsilon_1+\varepsilon_2)/\varepsilon kd}}\,,
\label{omegapkd}
\end{equation}
where $k$ is the in-plane momentum absolute value. As $d$ decreases, this $\omega_p(k)$ is seen to shift to the red, acquiring the $\sqrt{k}$ spatial dispersion of 2D materials~\cite{Basov}. With $d$ increasing it gradually approaches $\omega_p^{3D}$, the \emph{screened} bulk material plasma frequency. With this in mind, well below the interband transition frequencies, one can use the confinement-induced \emph{nonlocal} Drude dielectric function
\begin{equation}
\frac{\epsilon(k,\omega)}{\varepsilon}=1-\frac{\omega_p^{2}(k)}{\omega(\omega+i\gamma)}
\label{Lindhard}
\end{equation}
for the low-frequency response of the plasmonic film in the TD regime~\cite{bondmoushal18}. Here, $\gamma$ is the phenomenological inelastic electron scattering rate and $\omega_p(k)$ is given by Eq.~(\ref{omegapkd}). For metals at frequencies below the interband transition frequencies, $\varepsilon$ is known to be $\sim\!10$ in magnitude, being contributed both by the positive background of ions and by the interband transitions to some extent as well~\cite{mangar14,Lavrinenko19}.

Under continuous low-intensity light illumination the plasmonic film can still be treated as being at the thermal equilibrium. The thermal averaging of Eq.~(\ref{omegapkd}) then gives $\overline{\omega}_p(d,T)=\!\int_0^{k_c}\!dkk\,\omega_p(k)n(k)/\!\int_0^{k_c}\!dkkn(k)$. Here, the numerator sums up over all the plasma frequency modes with different $k$ that are occupied at temperature $T$ with the mean occupation number $n(k)=[e^{\,\hbar\omega_p(k)/k_BT}\!-1]^{-1}$, while the denominator provides the total number of such modes in the 2D $k$-space (bounded above by the 2D plasmon cutoff $k_c$).~This expression was recently used to explain a peculiar plasma frequency $T$-dependence observed for $100$~nm thick TiN films at cryogenic temperatures~\cite{Lavrinenko19}. At not too low $T$, for very thin films, $\overline{\omega}_p(d,T)$ loses its $T$ dependence to take the form
\begin{equation}
\overline{\omega}_p(d)=\frac{2C^2d^2\;\omega_p^{3D}}{(1+2Cd)\sqrt{Cd(1+Cd)}-\sinh^{-1}(\sqrt{Cd}\,)}
\label{omegapd}
\end{equation}
with $C=\varepsilon k_c/(\varepsilon_1+\varepsilon_2)$, to give the $\sqrt{d}$ thickness behavior in the ultrathin regime. This agrees well with the recent room-$T$ measurements~\cite{Shah17} and simulations~\cite{Shah18} of $\omega_p$ for stoichiometrically perfect TiN films of controlled variable thickness. Within its applicability domain, Eq.~(\ref{omegapd}) can be used to obtain $\varepsilon$ and/or $k_c$ with $C$ being treated as a parameter to fit experimental data in terms of the standard \emph{local} Drude model, which is typically the case in relevant experiments~\cite{Shah17,Lavrinenko19}.

\section{Transdimensional ENZ modes}

A straightforward way to elucidate the real nature of the ENZ modes of the ultrathin plasmonic films in the TD regime is to look at the dipolar spontaneous emission which in close proximity to the film is controlled by the near-field electromagnetic (EM) coupling to these modes. Consider an excited two-level atom (a point-like DE) positioned at $\textbf{r}_{A}\!=\!z_{A}\textbf{e}_z$ above the surface of the film as sketched in the inset of Fig.~\ref{fig3}~(a). In absence of external EM radiation, such an emitter couples to its surrounding vacuum EM field via the transition dipole moment $d_\mu\!=\!\langle u|\hat{d}_\mu|l\rangle$ ($\mu\!=\!x,y,z$) between the lower $|l\rangle$ and upper $|u\rangle$ atomic states separated by the frequency $\omega_{ul}$ (which we merely abbreviate as $\omega$ in what follows). For such a quantum system, the rigorous medium-assisted QED approach gives the spontaneous decay rate in the form~\cite{WelschQO,BuhmannPRA08} (Gaussian units)
\begin{equation}
\Gamma(z_A,\omega)=\Gamma_0+\frac{8\pi\omega^{2}}{\hbar c^{2}}\!\!\!\!\!\sum_{\mu,\nu=x,y,z}\!\!\!\!\!d_\mu d_\nu\,\mathrm{Im\,}G^{\mathrm{\,sc}}_{\mu\nu}(z_A,z_A,\omega),
\label{Gamma}
\end{equation}
where the first and second terms represent the free space and interface scattering contributions, respectively. Here, $G^{\mathrm{\,sc}}_{\mu\nu}(z,z^\prime,\omega)$ is the scattering part of the Green tensor of a planar multilayer structure [sketched in Fig.~\ref{fig3}~(a) in our case]~\cite{Tomas95}, which can be diagonalized to take the form~\cite{BuhmannPRA08}
\begin{eqnarray}
G^{\mathrm{\,sc}}_{\mu\mu}(z_{A},z_{A},\omega)=\!\frac{i}{2}\int^{+\infty}_{0}\!\!\!\!\!\!\!dk\frac{k}{\beta_2}\,R_{\mu\mu}(k)\,e^{2i\beta_{2}z_{A}},\hskip0.75cm\label{Greentensor}\\
R_{xx,yy}=r^{s}_{2-}(k)-\frac{\beta^{2}_{2}c^{2}}{\varepsilon_{2}\omega^{2}}\,r^{p}_{2-}(k),\;\;R_{zz}=\frac{2k^{2}c^{2}}{\varepsilon_{2}\omega^{2}}\,r^{p}_{2-}(k)\nonumber
\end{eqnarray}
with the integration done over the absolute value $k$ of the in-plane momentum component of spontaneously emitted $s$- and $p$-polarized photons (\emph{TE} and \emph{TM} waves, respectively). Their reflection coefficients are given by~\cite{Tomas95}
\begin{eqnarray}
r^{\sigma}_{2-}=\frac{r^{\sigma}_{2}-r^{\sigma}_{1}e^{2i\beta d}}{1-r^{\sigma}_{1}r^{\sigma}_{2}e^{2i\beta d}}\;\;\;(\sigma=s,p),\hskip1.2cm\label{rsigma}\\
r^{s}_{j}=\frac{\beta_{j}-\beta}{\beta_{j}+\beta}\,,\;\;\;r^{p}_{j}=\frac{\beta_{j}\,\epsilon(k,\omega)-\beta\varepsilon_{j}}{\beta_{j}\,\epsilon(k,\omega)+\beta\varepsilon_{j}}\;\;\;(j=1,2),\nonumber\\
\beta_{j}=\sqrt{\varepsilon_{j}k_0^2-k^{2}},\;\;\beta=\sqrt{\epsilon(k,\omega)k_0^2-k^{2}}\,.\hskip0.7cm\nonumber
\end{eqnarray}
Here, $\beta_{1,2}$ and $\beta$ are the absolute values of the photon momentum $z$-components in region~$1$ (substrate), in region~$2$ where the emitter is located, and in the film itself, respectively, $\epsilon(k,\omega)$ is the film response function of Eq.~(\ref{Lindhard}), and $k_0\!=\!\omega/c$.~Rescaling of the quantities in Eq.~(\ref{Greentensor}) by
\begin{equation}
z_A\!=\!\frac{l_{\!A}}{2\kappa},\;\;R_{\mu\mu}\!=\!\frac{2}{\kappa}\bar{R}_{\mu\mu},\;\;k\!=\!\kappa t,\;\;i\beta_2\!=\!\kappa x,\;\;\beta_2\!=\!\kappa y
\label{dimlessunits}
\end{equation}
with $\kappa\!=\!k_0\sqrt{\varepsilon_2}\,$, allows one to rewrite it as a sum of the two well-defined single-valued real integrals of the form
\begin{eqnarray}
G^{\mathrm{\,sc}}_{\mu\mu}(z_{A},z_{A},\omega)=
i\!\!\int^{+\infty}_{0}\!\!\!\!\!\!\!dt\frac{t}{\sqrt{1\!-\!t^2}}\,\bar{R}_{\mu\mu}(t)\,e^{i\sqrt{1-t^2}\,l_{\!A}}\label{propevan}\hskip0.5cm\\
=\!\!\int^{+\infty}_{0}\!\!\!\!\!\!\!\!\!dx\,\bar{R}_{\mu\mu}(\!\sqrt{1\!+\!x^2}\,)\,e^{-xl_{\!A}}\!
+i\!\!\int^{1}_{0}\!\!\!dy\,\bar{R}_{\mu\mu}(\!\sqrt{1\!-\!y^2}\,)\,e^{iyl_{\!A}}.\nonumber
\end{eqnarray}
Here, the term containing an exponentially damped factor results from evanescent waves and the term containing an oscillating factor results from propagating waves, to contribute the most near the surface and at large distances from the surface of the film, respectively.

\begin{figure}[t]
\includegraphics[scale=0.45]{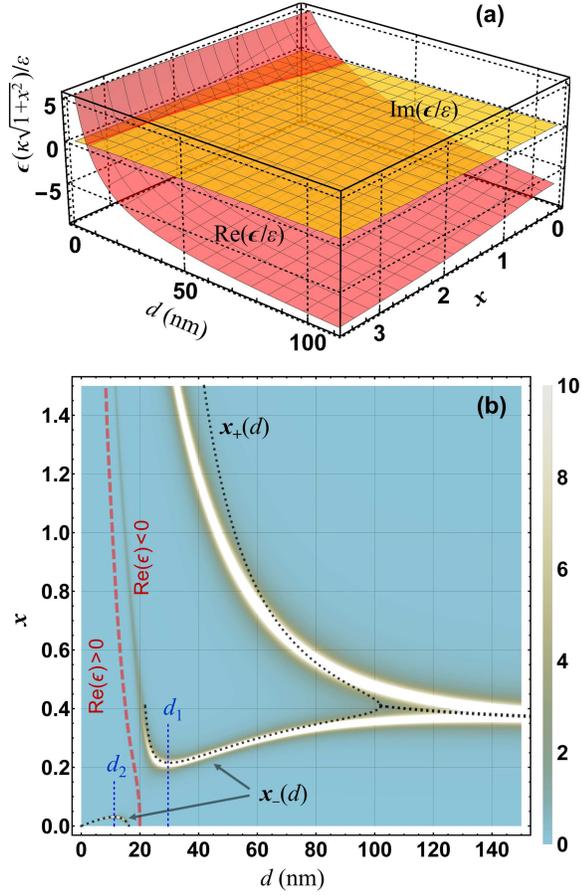}
\caption{(Color online) (a)~The confinement induced \emph{nonlocal} Drude dielectric response of Eq.~(\ref{epsilonnew}) as a function of $d$ and~$x$ (see text for other parameters chosen). (b)~The density plot of the $\mbox{Im}\,r^p_{2-}(\sqrt{1\!+\!x^2}\,)$ reflection coefficient as given by Eq.~(\ref{rsigma}). Black dotted lines show the real parts of the approximate modes of Eq.~(\ref{xpm}).}
\label{fig1}
\end{figure}

In general, Green's tensor poles on the real axis of the complex momentum space define the dispersion relations for the eigen modes of the problem~\cite{Abrikosov}. These poles in Eq.~(\ref{Greentensor}) come from the reflection coefficient poles. From Eq.~(\ref{rsigma}) one can see that only the $r^{p}_{2-}$ coefficient can have poles representing the ENZ modes of the ultrathin film, in which case $\epsilon(k,\omega)\!\approx\!0$ and $d\!\rightarrow\!0$. Further inspecting Eqs.~(\ref{rsigma})--(\ref{propevan}) one can find that only the $p$-evanescent wave coefficient $r^p_{2-}(\sqrt{1\!+\!x^2}\,)$ can have poles on the real axis while the $p$-propagating wave coefficient cannot. Indeed, since $\beta\!\approx\!ik$ for $\epsilon(k,\omega)\!\approx\!0$, the factor $e^{2i\beta d}\!\approx\!e^{-2kd}\!\sim\!0$ for all finite $d$ and sufficiently large $k$, yielding $r^p_{2-}\!\sim r^p_2$ which is clearly seen to have a pole on the real axis for evanescent waves only ($i\beta_2\!=\!\kappa x$) and not for propagating waves ($\beta_2\!=\!\kappa y$). Finally, only those poles of $r^p_{2-}(\sqrt{1\!+\!x^2}\,)$ located in the domain $0\!<\!x\!<\!1$ can significantly contribute to $G^{\mathrm{\,sc}}_{\mu\mu}(z_{A},z_{A},\omega)$ due to the presence of the exponential damping factor $e^{-xl_{\!A}}$ in the first integral of Eq.~(\ref{propevan}). With this in view and assuming $\varepsilon_1\!\approx\!\varepsilon_2$ for simplicity, the poles of interest come out as zeros of the denominator of the $r^p_{2-}(\sqrt{1\!+\!x^2}\,)$ coefficient. They can be obtained by expanding the denominator in Maclaurin series through the second order in $x$. The roots of the quadratic trinomial thus obtained give the two ENZ dispersion modes as follows
\begin{eqnarray}
x_{\pm}(d)=-b\pm\sqrt{b^2-c}\,,\;\;\;c(d)=\frac{\varepsilon_2/\epsilon-1}{\epsilon/\varepsilon_2+\varepsilon_2/\epsilon},\hskip1cm\label{xpm}\\
b(d)\!=\!\frac{\sqrt{\epsilon/\varepsilon_2\!-\!1}\cot(\kappa d\sqrt{\epsilon/\varepsilon_2\!-\!1}\,)}{\epsilon/\varepsilon_2+\varepsilon_2/\epsilon},\,\epsilon\!=\!\epsilon(\kappa\sqrt{1\!+\!x^2})\!\mid_{x=0}.
\nonumber
\end{eqnarray}
Here, as follows from Eqs.~(\ref{omegapkd}) and (\ref{Lindhard}) after rescaling (\ref{dimlessunits}),
\begin{equation}
\frac{\epsilon(\kappa\sqrt{1\!+\!x^2})}{\varepsilon}\!=\!1\!-\!\frac{\varepsilon k_p^{3D}d\sqrt{1\!+\!x^2}}{(2\sqrt{\varepsilon_2}+\varepsilon k_p^{3D}d\sqrt{1\!+\!x^2}\,u)(u\!+\!i\delta)}
\label{epsilonnew}
\end{equation}
with $u\!=\!\omega/\omega_p^{3D}$ and $\delta\!=\!\gamma/\omega_p^{3D}$ being the dimensionless analogues of their respective quantities, and $k_p^{3D}\!\!=\!\omega_p^{3D}\!/c$. This equation exhibits the explicit $d$ dependence coming from the confinement-induced \emph{nonlocal} Drude dielectric response of the KR model we use, which is seen to turn into the standard local Drude response as $d$ increases.

Figure~\ref{fig1}~(a) shows Eq.~(\ref{epsilonnew}) as a function of $d$ and~$x$ with other parameters chosen to take on moderate values typical of systems such as, for instance, nitrogen-vacancy centers in nanodiamonds near the TiN surface~\cite{plasmaTiN,SimeonBogdanov}. They are $\omega_p^{3D}\!=\!2.79$~eV~\cite{plasmaTiN}, $\varepsilon\!=\!7.8$~\cite{Lavrinenko19}, $\varepsilon_2\!=\!\varepsilon_1\!=\!1$, $u\!=\!0.65$~\cite{SimeonBogdanov}, and $\delta\!=\!0.01$. These parameters are used as representative examples for the figures shown in this work. The sharp $d$ dependence one can see in Fig.~\ref{fig1}~(a) in the domain where $\mbox{Re}\,\epsilon\!\approx\!0$ indicates that the ENZ modes of ultrathin films can be controlled by adjusting the film thickness. Such an opportunity disappears with the thickness increase, however, whereby $\mbox{Re}\,\epsilon$ becomes a large negative constant.

Figure~\ref{fig1}~(b) compares the ENZ modes visualized by the density plot of the imaginary part of the $r^p_{2-}(\sqrt{1\!+\!x^2}\,)$ reflection coefficient [contributing the most to Eq.~(\ref{Gamma}) as per Eq.~(\ref{propevan})] calculated directly from Eq.~(\ref{rsigma}), to the approximate modes of Eq.~(\ref{xpm}). The functions $\mbox{Re}\,x_\pm(d)$ are plotted by the black dotted lines. The correspondence is seen to be nearly perfect for $x\!<\!1$ where the approximation is valid. We see the splitting of the doubly degenerated mode with decreasing film thickness. As $d$ decreases, the higher momentum branch $x_+$ goes straight up and approaches the $\mbox{Re}\,\epsilon\!=\!0$ line [red dashed line, also highlighted in (a)] from the negative side. The lower momentum branch $x_-$ goes slowly down, passes through the minimum, rises up abruptly, and then dies away on the $\mbox{Re}\,\epsilon\!<\!0$ side. It then pops up on the $\mbox{Re}\,\epsilon\!>\!0$ side though, passes through the maximum, and tends to zero linearly with $d$ going down to zero. The analytical expressions of Eqs.~(\ref{xpm}) and (\ref{epsilonnew}) allow us to study these universal peculiar features (apparently originating from the \emph{nonlocal} dielectric response) for both modes, and thus to generalize the results reported earlier for only one of the modes (the one with higher momentum) within the local Drude response model~\cite{Brener15,Greffet19}. For example, while for conventional thin films $x_+(d)\!=\!x_-(d)\!=\!-b(d)$ since $\mbox{Re}\,\epsilon\!\ll\!-1$ and $b^2(d)\!=\!c(d)$ at large $d$, for ultrathin films in the TD regime one has $b^2(d)\!\gg\!c(d)$ and the degeneracy is lifted to give the two split modes as follows
\begin{equation}
x_+(d)\!=\!-\frac{2}{\kappa d\left(\epsilon/\varepsilon_2+\varepsilon_2/\epsilon\right)},\;\;x_-(d)\!=\!\frac{\kappa d}{2}\!\left(1-\frac{\varepsilon_2}{\epsilon}\right)\!.
\label{ultrathinlimits}
\end{equation}
Here, the upper mode $x_+(d)$ can be shown to reproduce the thickness dependence of the long-range plasmon dispersion reported earlier in Ref.~\cite{Brener15} within the \emph{local} Drude response model, which follows from our Eq.~(\ref{epsilonnew}) in the large $d$ limit. The lower mode $x_-(d)$ exhibits the peculiar features mentioned above. Testing it for extrema leads to $\,d_{1,2}\!=\!2\sqrt{\varepsilon_2}\,u\,[1\pm1/\!\sqrt{\varepsilon\!-\!(\varepsilon\!-\!1)u^2}\,]/[\varepsilon k_p^{3D}(1-u^2)]$ for the local minimum and maximum of the $x_-(d)$ function, respectively [traced by the vertical blue dashed lines in Fig.~\ref{fig1}~(b)]. This allows one to control these features by adjusting the film parameters appropriately.

\begin{figure}[t]
\includegraphics[scale=0.37]{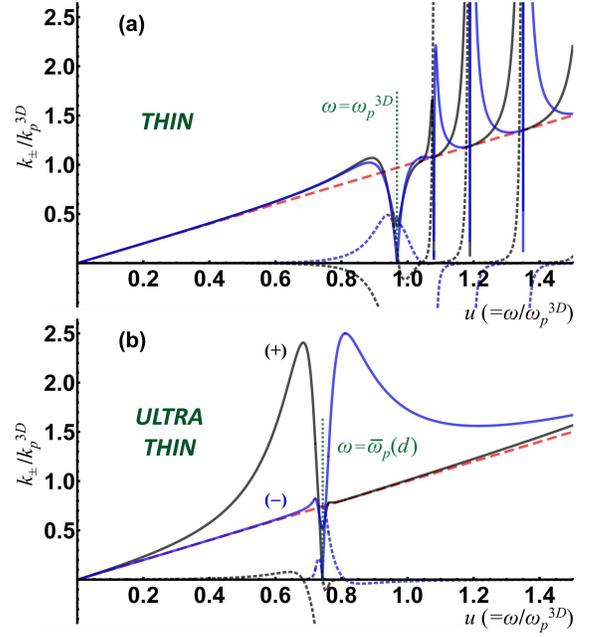}
\caption{(Color online) The dispersion relations $k(\omega)$ of the SP modes for (a) $300$~nm and (b) $30$~nm thick films as given by Eq.~(\ref{SPmodes}). Short-range ($-$) and long-range ($+$) modes are shown by the blue and black lines, respectively, with solid and dashed lines tracing their respective real and imaginary parts (only positive imaginary part solutions must be accepted for causality reasons). Dashed red line is the light cone line.}
\label{fig2}
\end{figure}

From Eq.~(\ref{xpm}) one can also obtain explicitly the dispersion relations for the short- and long-range SP modes (within the limits of our approximation) as given by the KR model \emph{nonlocal} dielectric response (\ref{epsilonnew}) that we use here. With the relation between $k$ and $x$ provided by Eq.~(\ref{dimlessunits}), one obtains the two SP modes as follows
\begin{equation}
k_{\pm}=\kappa\sqrt{1+x_{\pm}^2}=k_p^{3D}\sqrt{\varepsilon_2(1+x_{\pm}^2)}\,u
\label{SPmodes}
\end{equation}
with $x_{\pm}(d)$ given by Eq.~(\ref{xpm}), or by Eq.~(\ref{ultrathinlimits}) for the ultra\-thin film case in the TD regime. These are plotted in Fig.~\ref{fig2} for two different thicknesses to demonstrate the effect of thickness on the short- and long-range SP modes. They are $d\!=\!300$~nm in (a) and $d\!=\!30$~nm in (b) [cf. Fig.~\ref{fig1}~(b)], with all other parameters being the same. Highlighted in blue and black are the short- and long-range SP dispersion modes, respectively, with the solid and dashed lines representing their respective real and imaginary parts. The dashed red line depicts the light cone line. Taking the square root in Eq.~(\ref{SPmodes}) generates solutions with both positive and negative imaginary parts. However, only the positive imaginary part solutions must be accepted for causality reasons, according to how the scattering Green tensor in Eq.~(\ref{Greentensor}) is defined in our case, which assumes the $e^{-i(\omega t-\textbf{k}\cdot\textbf{r})}$ space-time convention.

In general, the momentum-frequency dispersion rela\-tion can be rigorously defined mathematically in two equivalent ways~\cite{TeperikPRB09}. They are either using a real-valued $\omega$ to determine the complex-valued $k$, or using a real-valued $k$ to determine the complex-valued $\omega$. Figure~\ref{fig2} presents the dispersion relations in the form $k(\omega)$ with the real-valued $\omega$ and the complex-valued in-plane $k$, where the actual eigen modes of the system are given by the graph segments with the positive imaginary parts of $k$ as mentioned above. One can clearly see a big difference between the conventional thin films and the ultrathin TD films of the same material composition. In Fig.~\ref{fig2}~(a), the two SP modes are degenerated below the plasma frequency. Above the plasma frequency they turn into half-wavelength modes responsible for the resonance light reflection (or transmission) as expected~\cite{BornWolf}. The plasma frequency itself (traced by the vertical green dotted line) is very close to $\omega_p^{3D}$ and $d$ independent. In Fig.~\ref{fig2}~(b), the mode degeneracy is lifted below the plasma frequency, and the half-wavelength modes disappear to reveal a new feature in a very narrow frequency range above the plasma frequency. Here, the short-range (blue) SP mode turns into a very sharp [flat if inverted to the $\omega(k)$ form] long-range SP mode. This overall effect looks similar to that earlier reported for ultra\-thin films within the standard local Drude response model~\cite{Brener15}. However, the KR confinement-induced \emph{nonlocal} response model we use herein reveals new and essential details. Firstly, at fixed $d$ the effect occurs when $\mbox{Re}\,\epsilon$ of Eq.~(\ref{epsilonnew}) changes its sign from negative to positive as $\omega$ increases. Secondly, the SP dispersion in this domain can be found analytically from Eqs.~(\ref{ultrathinlimits}) and (\ref{SPmodes}). Finally and most importantly, the effect can be controlled by adjusting the parameters of the ultrathin TD film such as $\varepsilon$, $\varepsilon_{1,2}$ and $d$ through its plasma frequency which according to Eqs.~(\ref{omegapkd}) and~(\ref{omegapd}) is now thickness dependent.

\begin{figure}[t]
\includegraphics[scale=0.37]{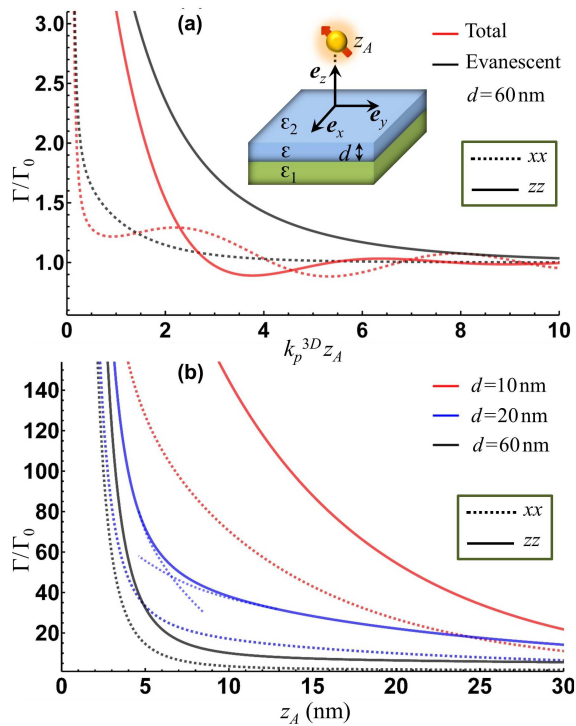}
\caption{(Color online) (a)~Dimensionless distance dependence of the $\Gamma/\Gamma_0$ ratio of Eq.~(\ref{Gamma}) for the perpendicular ($zz$) and parallel ($xx$) DE orientation (sketched in the inset). (b)~Same for the distance $z_A$ in nanometers, with the largest contribution coming from the evanescent term in Eq.~(\ref{propevan}). The dotted blue line segments guide the eye for the biexponential decay.}
\label{fig3}
\end{figure}

\section{Biexponential distance dependence}

For ultrathin TD plasmonic films, due to their SP mode degeneracy lifting, a DE in close proximity to the surface of the film can couple to each of the split ENZ modes individually. Figure~\ref{fig3} presents the spontaneous decay rates relative to vacuum ($\Gamma_0\!=\!4|\textbf{d}|^2\omega^3/3\hbar c^3$~\cite{WelschQO}) calculated from Eq.~(\ref{Gamma}) using Eqs.~(\ref{Greentensor})--(\ref{propevan}).~In Fig.~\ref{fig3}~(a), one can see the dimensionless distance dependence for the perpendicular ($zz$) and parallel ($xx$) dipole orientation near the film of thickness $d\!=\!60$~nm as sketched in the inset. The evanescent wave contribution is shown separately and is seen to be determinative, whereas the propagating wave contribution gives the oscillatory distance-dependent behavior. At this thickness, from Fig.~\ref{fig1}~(b) one can find the two ENZ modes, $x_+(d)$ and $x_-(d)$, that are available for the DE to couple to, thus leading to a peculiar thickness-controlled feature --- the biexponential distance dependence of the spontaneous decay rate. This can be seen in details in Fig.~\ref{fig3}~(b), which presents decay rates as functions of the DE-surface distance in nanometers for thicknesses $d\!=\!10$, $20$, and $60$~nm. Here, the evanescent wave contribution totally dominates, and the overall decay rate enhancement is between two and three orders of magnitude. The dotted blue line segments guide the eye for the biexponential distance dependence effect, whereby the DE-surface distance dependence of the spontaneous decay enhancement factor is simultaneously controlled by the two split ENZ modes, $x_+(d)$ and $x_-(d)$, of Eq.~(\ref{ultrathinlimits}). Indeed, taking advantage of the sharp peak structure of the $\mbox{Im}\,r^p_{2-}(\sqrt{1\!+\!x^2}\,)$ reflection coefficient shown in Fig.~\ref{fig1}~(b) and the fact that it contributes the most to the imaginary part of the evanescent term of the scattering Green tensor (\ref{propevan}) as discussed in the previous Section, one can use the Lorentzian approximation in Eq.~(\ref{propevan}) to write
\begin{equation}
\mbox{Im}\,\bar{R}_{\mu\mu}(\!\sqrt{1\!+\!x^2})\approx\frac{1}{\pi}\sum_{+,-}
\frac{\bar{R}_{\mu\mu}^{(\pm)}e^{-\mathrm{Re\,}x_{_{\!\pm}}l_{\!A}}\mathrm{Im\,}x_{\pm}}{(x\!-\!\mathrm{Re\,}x_{\pm})^2+(\mathrm{Im\,}x_{\pm})^2}
\label{Lorentzian}
\end{equation}
with $\bar{R}_{\mu\mu}^{(\pm)}\!=\mbox{Im\,}\bar{R}_{\mu\mu}[\sqrt{1\!+\!(\mathrm{Re\,}x_{\pm})^2}\,]$ and $\mathrm{Im\,}x_\pm$ being the half-width-at-half-maxima (representative of the modal damping) of the two normalized Lorentzian functions corresponding to the two resonances in Fig.~\ref{fig1}~(b). Now using Eq.~(\ref{Lorentzian}), one can easily integrate the evanescent term in Eq.~(\ref{propevan}) to obtain
\begin{eqnarray}
\mbox{Im}\,G^{\mathrm{\,sc(ev)}}_{\mu\mu}(z_{A},z_{A},\omega)=\!\int^{+\infty}_{0}\!\!\!\!\!\!\!\!\!dx\,\mbox{Im}\bar{R}_{\mu\mu}(\!\sqrt{1\!+\!x^2}\,)\,e^{-xl_{\!A}}\nonumber\\
\approx\frac{1}{\pi}\sum_{+,-}\bar{R}_{\mu\mu}^{(\pm)}e^{-\mathrm{Re\,}x_{_{\!\pm}}l_{\!A}}\!
\left[\arctan\!\left(\frac{\mathrm{Re\,}x_{\pm}}{\mathrm{Im\,}x_{\pm}}\right)+\frac{\pi}{2}\right]\hskip0.25cm\nonumber\\
\approx\bar{R}_{\mu\mu}^{(+)}e^{4(\mathrm{Re\,}\mbox{$\epsilon$}/d)z_{\!A}}+\bar{R}_{\mu\mu}^{(-)}e^{\kappa^{2}\!d(1/\mathrm{Re\,}\mbox{$\epsilon$}-1)z_{\!A}}.\hskip0.5cm
\label{biexponential}
\end{eqnarray}
Here, after the integration over $x$ the $\arctan$ function Taylor series expansion is done to the first nonvanishing order in $\mathrm{Im\,}x_{\pm}/\mathrm{Re\,}x_{\pm}\!\ll\!1$, followed by using the explicit expressions for $x_{\pm}(d)$ as given by Eq.~(\ref{ultrathinlimits}) with $\varepsilon_2\!=\!1$ and $\epsilon\!\rightarrow\!0$ as prescribed by Eq.~(\ref{epsilonnew}) and shown in Fig.~\ref{fig1}~(a).

Equation~(\ref{biexponential}) presents a determinative contribution to the DE spontaneous decay rate enhancement factor in Eq.~(\ref{Gamma}). This is the sum of the two terms with $z_{\!A}$-dependent exponential damping factors that are opposite in their $d$-dependences. The two exponentials are defined in the domains $\mathrm{Re\,}\epsilon\!<\!0$ and $\mathrm{Re\,}\epsilon\!<\!0\cup\mathrm{Re\,}\epsilon\!>\!1$, respectively, as discussed for Fig.~\ref{fig1}~(b) in the previous Section. At sufficiently small $d$, both of them make the DE spontaneous decay enhancement factor decrease as $z_{\!A}$ increases moving the DE away from the film surface. However, the argument of the first exponential increases in absolute value with decreasing $d$, which makes it only significant for short-range $z_{\!A}$ near the surface of the film. The argument of the second exponential decreases with $d$ in absolute value, which makes it significant both for short-range and for long-range $z_{\!A}$ in Eq.~(\ref{Gamma}). It is this exponential that holds the decay enhancement factor large in magnitude over DE-surface distances much longer than those controlled by the first exponential alone.

Decreasing $d$ to transition from the $\mathrm{Re\,}\epsilon\!<\!0$ domain to the $\mbox{Re}\,\epsilon\!>\!1$ domain leaves only the second exponential of a tiny negative argument in Eq.~(\ref{biexponential}). Only the $x_-$ mode with $x\lesssim0.01$ is available there as one can see from Fig.~\ref{fig1}~(b), which is a remarkable feature of the confinement-induced nonlocality. It is the DE coupling to this mode that pushes up the spontaneous decay ratio in Eq.~(\ref{Gamma}) to give over three orders of magnitude enhancement shown in Fig.~\ref{fig3}~(b) for $d\!=\!10$~nm. Lowering $d$ even further down makes $\epsilon\!=\!\varepsilon$ in the limit $d\!\rightarrow\!0$ as prescribed by Eq.~(\ref{epsilonnew}). Microscopically, this limit does not make sense as $\varepsilon$ can only be introduced for a \emph{physically} small material volume. However, since the reflection coefficients in Eq.~(\ref{rsigma}) are obtained from macroscopic boundary conditions, they do remain well defined for $d\!=\!0$ as well, in which case they describe the interface light scattering rather than the scattering by the film. Therefore, in this limit (and only in this limit) our results transition into the standard results of macroscopic surface optics~\cite{BornWolf}.

The stronger $z$-oriented dipole decay comes from the dipole emission angular distribution being predominantly perpendicular to the axis of the dipole. This can also be understood from the mirror charge corollary of the electrostatic uniqueness theorem, whereby an electric dipole oriented perpendicular to a conducting plane generates a collinear mirror image dipole, while the same dipole oriented parallel to the plane generates the anti-collinear image dipole. The total dipole moment (original+image) is larger in the former than in the latter case, leading to the greater spontaneous decay enhancement factor for the DE orientation perpendicular to the film surface.

\section{Concluding remarks}

In this article, we use macroscopic QED and the confinement-induced nonlocal Drude dielectric response model based on the Keldysh-Rytova pairwise electron interaction potential to study the ENZ modes and their coupling to a point-like atomic dipole emitter in close proximity to the surface of an ultrathin plasmonic film in the transdimentional regime. As opposed to the conventional thin film models studied previously that rely on either purely 2D material properties, or on 3D materials with macroscopic boundary conditions imposed on their top and bottom interfaces~\cite{Ritchie57,Economou69,DahlSham77,Theis80,AndoFowlerStern82,Chaplik85,Wang96,Pitarke07,Politano14}, the KR model we use herewith takes explicitly into account the vertical confinement effects in the ultrathin TD films.

Our results generalize the fundamental work by Drexhage~\cite{Drexhage} as well as those of related recent works~\cite{Brener15,Greffet19} by specifically demonstrating how the light-matter interaction properties in finite-thickness metallic films evolve with their thickness decrease from the bulk material properties to those of 2D plasmonic materials. We report new remarkable thickness-controlled effects for the ultrathin plasmonic TD films. They are the SP mode degeneracy lifting and the DE coupling to the ENZ modes split. This coupling leads to the biexponential DE-surface distance dependence of the spontaneous decay with rates two-to-three orders of magnitude greater than in free space. Importantly, these effects can be controlled due to the thickness-dependent plasma frequency of the TD metallic film~\cite{bondshal17} --- a unique microscopic property that cannot be obtained from the macroscopic boundary conditions imposed on the bulk metal film interfaces. The vertical electron confinement turns the electron-electron Coulomb potential into the much stronger and thickness-dependent KR interaction potential~\cite{Keldysh79}, leading to the thickness-dependent plasma oscillation frequency of the TD film and thus providing the possibility to control the light-matter interactions, the magneto-optical response, and the near-field properties of the ultrathin metallic films in the TD regime~\cite{bondshal17,bondmoushal18,bond19}.

Knowledge of these features is advantageous both for the fundamental understanding of electromagnetic properties and for the development of the new design principles of efficient photonic nanodevices with desired characteristics that are built on ultrathin plasmonic TD films.

\acknowledgments

This research is supported by the U.S. National Science Foundation under Condensed Matter Theory Program Award \# DMR-1830874 (I.V.B.) and by the U.S. Department of Energy, Office of Basic Energy Sciences, Division of Materials Sciences and Engineering under Award \# DE-SC0017717 (V.M.S.). H.M. was funded from the DOE BES grant \# DE-SC0007117 awarded to I.V.B.

\end{document}